\documentstyle{article}
\begin{document}
\begin{center}
{\Large {\bf Extended Version of \\
``The Philosophy of the Trajectory Representation \\
of Quantum Mechanics''}} \\

\bigskip

Edward R. Floyd \\

10 Jamaica Village Road, Coronado, California 92118-3208, USA \\

floyd@crash.cts.com \\

\bigskip

\bigskip

\begin{minipage}{5.5in}

\noindent {\small  The philosophy of the trajectory representation differs with 
Copenhagen and Bohmian philosophies. The trajectory representation is a strongly 
causal, nonlocal theory of quantum mechanics that is deterministic.  It is 
couched in a generalized Hamilton-Jacobi formulation. For bound states, each 
particular trajectory determines a unique microstate of the Schr\"{o}dinger wave 
function. Hence, the Schr\"{o}dinger wave function is not an exhaustive 
description of nonrelativistic quantum phenomenon.  A tunnelling example shows 
that assigning a probability amplitude to the Schr\"{o}dinger wave function is 
unnecessary.  The trajectory representation in the classical limit ($\hbar \to 
0$) manifests a residual indeterminacy where the trajectory representation does 
not go to classical mechanics.  This residual indeterminacy is contrasted to the 
Heisenberg uncertainty principle and is also compared with 't~Hooft's 
information loss.  The trajectory representation is contrasted with the 
Copenhagen and Bohmian representations.  For a square well duct, consistent 
overdetermination of a trajectory by a redundant set of observed constants of 
the motion are beyond the Copenhagen interpretation.  Also, the trajectory 
representation makes different predictions than the Copenhagen interpretation 
for impulsive perturbations, even under Copenhagen epistemology.  Although the 
trajectory representation and Bohmian mechanics use the same generalized 
Hamilton-Jacobi equations, they have different equations of motion.}  

\end{minipage}

\end{center}

\paragraph{Prologue:}  ``The Philosophy of the Trajectory Representation of 
Quantum Mechanics" [\ref{bib:f}] is an extract of this opus, was presented at 
the Vigier 2000 Symposium, 21--25~August~2000, in Berkeley, California, and will 
be published in the {\it Proceedings} of the Symposium.  The Symposium 
celebrated Jean Pierre Vigier's eightieth birthday.   

\paragraph{1. Introduction:}  The seminal work on the trajectory representation 
was published in 1982 [\ref{bib:f82b}].  The trajectory representation sprang 
from improvements in the WKB approximation [\ref{bib:f76a}--\ref{bib:f82a}] and 
in acoustical
ray tracing [\ref{bib:f76b},\ref{bib:f86b}].  The equations of motion for the 
trajectories are developed from a quantum Hamilton-Jacobi formulation.  These 
trajectories are deterministic and continuous.  Ergo, there is no need by 
precept for any collapse of the wave function during observation.   Early 
analyses used numerical methods or power-series expansions until exact 
closed-form
solutions were introduced [\ref{bib:f86c}].

Recently, Faraggi and Matone [\ref{bib:fm98a}--\ref{bib:fm00}] have 
independently generated the same quantum Hamilton-Jacobi formulation from an 
equivalence principle free from any axioms.  Faraggi and Matone have shown that 
although all quantum systems can be connected by an equivalence coordinate 
transformation (trivializing map), all systems in classical mechanics are not so 
connected.  Some of the goals of their work include synthesis of gravity, mass 
and quantum mechanics and possible relations to string theory 
[\ref{bib:fa00a},\ref{bib:fa00b}] and producing an expression for the 
interaction terms, including gravity, that have a pure quantum origin 
[\ref{bib:ma00}].  The development of the equivalence principle is beyond the 
scope of this exposition.      

We present the philosophical aspects of the trajectory representation of
quantum mechanics that distinguish this representation.  We exhibit its 
interpretation, which we contrast to the Copenhagen interpretation and the 
Bohmian stochastic interpretation.  Our findings are presented in closed form in
one dimension for the time-independent case whenever one dimension suffices.  
The work in one dimension for the time-independent case renders a
counter example that refutes Born's postulate of the Copenhagen
interpretation attributing a probability amplitude to the
Schr\"{o}dinger wave function, shows that the Heisenberg uncertainty principle 
is premature, refutes the Copenhagen
interpretation that the Schr\"{o}dinger wave function is an
exhaustive description of nonrelativistic quantum phenomenon, and questions the 
wave-particle duality of Bohr's complementarity.  Bertoldi, Faraggi and Matone 
have recently extended the quantum Hamilton-Jacobi formulation to higher 
dimensions, time dependence and relativistic quantum mechanics 
[\ref{bib:bfm99}].  A small amount of work in higher dimensions is presented 
where necessary to establish our findings.

We explicitly note that the trajectory representation is not just another 
interpretation of quantum mechanics because it also predicts results that differ 
with contemporary, orthodox practice (Copenhagen interpretation).  Trajectory 
and Copenhagen analyses predict different results from a perturbing impulse 
[\ref{bib:f99}].  A test has been proposed to show that consistent 
overdetermination of a trajectory by a redundant set of observed constants of 
the motion would be beyond the Copenhagen interpretation [\ref{bib:f00b}].

Beyond the philosophical aspects, we refer the interested reader to
five other advances of the trajectory representation that have 
been developed elsewhere but not presented in the {\it Proceedings} 
[\ref{bib:f}].  First, an initial application of the trajectory
representation has been made to relativistic quantum mechanics
[\ref{bib:f88}].  Second, the trajectory representation is not a
hydrodynamical formulation of wave mechanics as trajectories may
cross. Thus, the trajectory representation may manifest caustics as
has been presented elsewhere, albeit couched in acoustics [\ref{bib:f86a}].  We 
note that the trajectory representation renders not
only all caustics that correspond to the caustics described by
classical ray tracing but also additional caustics that are extra
to classical geometric acoustics.  Third, creation and annihilation
of interference patterns are studied [\ref{bib:f94}].  Fourth,
trajectory dwell times during tunneling and reflection are examined 
[\ref{bib:f00b},\ref{bib:f95}].  Fifth, the generalized 
Hamilton-Jacobi equation and the Schr\"{o}dinger equation form an
Ermakov system which generates an Ermakov invariant [\ref{bib:f96a}]. 
The Ermakov invariant is a constant of the motion for the
particular trajectory (microstate).

In Section 2, we present the fundamentals of the trajectory
representation from a philosophical aspect.  We give references for more 
detailed development of the trajectory representation for the interested reader.  
The equations of motion are presented for the trajectory.  We present why 
microstates of the wave function exist for bound states.  Much of the
philosophy of the trajectory representation is innate in the
development of this representation.  In Section 3, we present different 
predictions rendered by trajectories and Copenhagen.   We continue to contrast 
in Section 4 the trajectory representation to the Copenhagen interpretation.   
In Section 5, we compare the trajectory representation with the
Bohmian stochastic representation.  In Appendix A, we show that no particular 
set of independent solutions of the Schr\"{o}dinger equation are privileged.       

\paragraph{2. The Trajectory Representation:}

\subparagraph{2.1.  Equation of Motion:} The trajectory representation is
based upon a phenomenological, nonlocal generalized 
Hamilton-Jacobi formulation.  The quantum stationary 
Hamilton-Jacobi equation (QSHJE) is given in one
dimension $x$ by [\ref{bib:f84},\ref{bib:f96b}] 

\begin{equation}
\frac{(W')^2}{2m}+V-E=-\frac{\hbar ^2}{4m}\langle W;x \rangle 
\label{eq:hje}
\end{equation}

\noindent where $W$ is Hamilton's characteristic function (also known as the 
reduced action), $W'$ is the momentum conjugate to $x$, $\langle W;x \rangle $ 
is the Schwarzian derivative of $W$ with respect to $x$, $V$ is the potential, 
$E$ is energy, $m$ is the mass of the particle, and $\hbar =h/(2\pi)$ where in 
turn $h$ is Planck's constant.  Explicitly, the Schwarzian derivative raises the 
QSHJE to a third-order nonlinear differential equation and is given by

\[
\langle W;x \rangle = \frac{W'''}{W'}-
\frac{3}{2}\left(\frac{W''}{W'}\right)^2 = 
[\ln (W')]''-\frac{1}{2}\{ [\ln (W')]'\} ^2. 
\]

\noindent The left side of Eq.\ (\ref{eq:hje})
manifests the classical Hamilton-Jacobi equation; the right side,
the higher order quantum effects in the Schwarzian derivative.  Faraggi and 
Matone have independently derived the QSHJE from the equivalence principle.  We 
note that $W$ and $W'$ are real even in a classically forbidden region.  The 
general solution for $W'$ is given by [\ref{bib:f86c}] 

\begin{equation}
W'= (2m)^{1/2}(a\phi ^2+b\theta ^2+c\phi \theta )^{-1}
\label{eq:cme}
\end{equation}

\noindent where $(a,b,c)$ is a set of real coefficients such that 
$a,b > 0$, and  $(\phi,\theta)$ is a set of normalized independent
solutions of the associated stationary Schr\"{o}dinger
equation, $-\hbar ^2\psi''/(2m) + (V-E)\psi = 0$.  The independent
solutions $(\phi,\theta)$ are normalized so that their Wronskian,
${\cal W}(\phi,\theta) = \phi \theta ' - \phi '\theta $, is scaled
to give  ${\cal W}^2(\phi,\theta) = 2m/[\hbar ^2(ab-c^2/4)] > 0$. 
This ensures that $(a\phi ^2 + b\theta ^2 + c\phi \theta) > 0$ and that $W'$ is 
real in the classically forbidden regions ($V>E$).  This normalization is 
determined by the nonlinearity of Eq.\ (\ref{eq:hje}) rather than by total 
probability of finding the particle in space be unity  as done by the Copenhagen 
interpretation.  A particular set $(\phi ,\theta )$ of independent solutions of 
the Schr\"{o}dinger equation may be chosen by the superposition
principle so that the coefficient $c$ is zero.  The motion in phase
space is specified by Eq.\ (\ref{eq:cme}).  This phase-space
trajectory is a function of the set of coefficients $(a,b,c)$.    

If the Schr\"{o}dinger equation can be solved in closed form, then the QSHJE may 
also be solved in closed form for conjugate momentum as Eq.\ (\ref{eq:cme}) 
expresses $W'$ in terms of products of $\phi $ and $\theta $.

In general, the conjugate momentum expressed by Eq.\ (\ref{eq:cme}) is not the 
mechanical momentum, i.e., $W' \ne m\dot{x}$.  Actually, $m\dot{x} = m \partial 
E/\partial W'$ [\ref{bib:f82b},\ref{bib:fm00}].

The solution for the generalized Hamilton's characteristic
function, $W$, is given by

\begin{equation}
W=\hbar \arctan \left(\frac{b(\theta /\phi ) + c/2}{(ab-
c^2/4)^{1/2}}\right)+K
\label{eq:hcfe}
\end{equation}

\noindent where $K$ is an integration constant that we may set to
zero herein.  

Hamilton's characteristic function is a generator of motion.  The
equation of motion in the domain $[x,t]$ is rendered by the
Hamilton-Jacobi transformation equation for constant coordinates
(often called Jacobi's theorem).  The procedure simplifies for
coordinates whose conjugate momenta are separation constants. Carroll has shown 
that for stationarity Jacobi's theorem is valid for $W$ is a Legendre transform 
of Hamilton's principal function [\ref{bib:ca99}]. For stationarity, $E$ is a 
separation constant for time.  Thus, the
equation of motion for the trajectory time, $t$, relative to its
constant coordinate $\tau $, is given as a function of $x$ by 

\begin{equation}
t-\tau= \partial W/\partial E
\label{eq:eom}
\end{equation}

\noindent where the trajectory is a function of a set of
coefficients $(a,b,c)$ and $\tau $ specifies the epoch.

The set $(\phi ,\theta)$ can only be a set of independent solutions
of the Schr\"{o}dinger equation.  Direct substitution of Eq.\
(\ref{eq:cme}) for $W'$ into Eq.\ (\ref{eq:hje}) gives

\begin{eqnarray}
&  & \frac{a\phi +c\theta /2}{a\phi ^2+b\theta ^2+c\phi \theta }[-
\hbar ^2/(2m) \phi ''-(E-V)\phi ] \nonumber \\
&  & + \frac{b\theta +c\phi /2}{a\phi ^2+b\theta ^2+c\phi \theta
}[-\hbar ^2/(2m) \theta ''-(E-V)\theta ] \nonumber \\
&  & \ \ \ \ \ \ \ \ \ \ \ \ \ \ \ \ \ \ - \frac{[{\cal W}^2\hbar
^2(ab-c^2/4)/(2m)-1]}{(a\phi ^2+b\theta ^2+c\phi \theta )^2} \  =
\  0. \label{eq:hjse}
\end{eqnarray}

\noindent For the general solution for $W'$, the real coefficients
$(a,b,c)$ are arbitrary within the limitations that  $a,b > 0$  and
from the Wronskian that $ab - c^2/4 > 0$.  Hence, for generality
the expressions within each of the three square brackets on the
left side of Eq.\ (\ref{eq:hjse}) must vanish identically.  The
expressions within the first two of these square brackets manifest
the Schr\"{o}dinger equation, so the expressions within these two
square brackets are identically zero if and only if $\phi $ and
$\theta$ are solutions of the Schr\"{o}dinger equation.  The
expression within third bracket vanishes identically if and only if
the normalization of the Wronskian is such that  ${\cal
W}^2(\phi ,\theta ) = 2m/[\hbar ^2(ab-c^2/4)]$.  For  ${\cal
W}(\phi ,\theta ) \neq 0$,  $\phi $ and $\theta $ must be independent
solutions of the Schr\"{o}dinger equation.  Hence, $\phi $ and
$\theta $ must form a set of independent solutions of the
Schr\"{o}dinger equation.

Equation (\ref{eq:hjse}) is independent of any particular choice of {\it 
ansatz}.  When comparing trajectories to Copenhagen and Bohm, we have broad 
selection for choosing a convenient {\it ansatz} to generate the equivalent wave 
picture (nothing herein implies that the trajectories need waves for 
completeness; only convenience).

\subparagraph{2.2.  Tunneling with Certainty:}  The Hamilton's characteristic 
function for the trajectory of a particle with sub-barrier energy that tunnels 
through the barrier with certainty can be established by the continuity 
conditions of $W,\ W'$ and $W''$ across the barrier interfaces [\ref{bib:f95}].  
The corresponding Schr\"{o}dinger wave function for this trajectory that tunnels 
with certainty was also developed from $W$ and $W'$ [\ref{bib:f95}].  We
now outline this development.

While Eq.\ (\ref{eq:cme}) gives the relationship between the
conjugate momentum $W'$ and the solution set of independent wave
functions $(\phi ,\theta )$, an inverse relationship, not
necessarily unique, is given by Ref.\ \ref{bib:f86c} as  

\begin{equation}
\psi  = \frac{\exp (iW/\hbar )}{(W')^{1/2}}. 
\label{eq:ansatz}
\end{equation} 

Let us consider a rectangular barrier whose potential is given by

\[
V(x) = \left\{ \begin{array}{cc}
            U, & \  \  |x|<q \\ [.1in]
            0, & \  \  |x|\geq q.
            \end{array}
       \right.
\]

\noindent For $x>q$, we specify a transmitted, unmodulated running
wave given by 

\begin{equation}
\psi  = (\hbar k)^{-1/2} \exp [ik(x-q)],\  \      x>q
\label{eq:tpsi}
\end{equation}

\noindent where $k=(2mE)^{1/2}/\hbar $ and the integration
constant, $K$, has been chosen so the phase is zero at the barrier
interface $x=q$.  In turn, $E$ is positive, sub-barrier, that is $0<E<U$.  For 
$x>q$, Hamilton's characteristic function is
given by $W=\hbar k(x-q)$.  Anywhere that $x>q$, $W=\hbar k(x-q)$ and its first 
two derivatives render a valid set of initial conditions. 

From the continuity of $W,\ W'$ and $W''$, we may now establish $W$ for this 
tunneling problem to be [\ref{bib:f95}]

\[
W = \left\{ \begin{array}{cc} \\
     		\hbar k(x-q),\ \ & \ \ x>q \\ [.1in]
		\hbar \arctan \{ (k/\kappa )\tanh[\kappa (x-q)\} ,
\ \ & \ \ -q\le x \le q \\ [.1in]
		\hbar \arctan ({\cal N}/{\cal D}), \ \ & \ \ x<q 
		\end{array}
	\right.\
\]

\noindent where $\kappa = [2m(U-E)]^{1/2}/\hbar $,

\[
{\cal N} = (k/\kappa )\sinh (-2\kappa q)\cos [k(x+q)]+\cosh (-2\kappa q)\sin 
[k(x+q)],
\]

\noindent and

\[
{\cal D} = \cosh (-2\kappa q) \cos[k(x+q)] + (\kappa /k)\sinh (-2\kappa q)\sin 
[k(x+q)].
\]

\noindent Note that $W$ monotonically increases everywhere with increasing $x$. 
While $W$, as given above, resolves tunneling in trajectory representation, we 
present the more familiar $\psi $ as derived from $W$ and Eq. (\ref{eq:ansatz}) 
to give a gentler introduction to the insights of the trajectory represtation.   

In the classically forbidden region inside the barrier, $-q \leq x \leq q$, and 
from Eq.\ (\ref{eq:ansatz}) the continuity conditions on $W,\ W'$ and
$W''$ at $x=q$, the Schr\"{o}dinger wave function is [\ref{bib:f95}]

\begin{equation}
\psi = \{ [(\kappa /k) \cosh ^2(\kappa x) + (k/\kappa ) \sinh
^2(\kappa x)]/(\hbar \kappa )\} ^{1/2} \exp \left [i \arctan
\left(\frac{k}{\kappa} \tanh [\kappa(x-q)]\right)\right],\  \   -q
\leq x \leq q
\label{eq:fpsi}
\end{equation}

\noindent where for Eqs.\ (\ref{eq:cme}) and (\ref{eq:hcfe}) $\phi = 
\cosh[\kappa (x-q)],\  \theta =\sinh[\kappa (x-q)],\  a = [(2m)^{1/2}/(\hbar 
\kappa
)](\kappa /k),\  b = [(2m)^{1/2}/(\hbar \kappa )](k/\kappa )$, and
$c=0$.  This Schr\"{o}dinger wave function represented by Eqs.\
(\ref{eq:tpsi}) and (\ref{eq:fpsi}) has a continuous logarithmic
derivative across the barrier interface at $x=q$.  The phase of $\psi $ inside 
the barrier increases monotonically with increasing $x$.  As Eq.\
(\ref{eq:fpsi}) manifests a spatially compound wave running in the
positive $x$-direction in the classically forbidden region that has
a continuous logarithmic derivative at $x=q$ with a wave that is
running in the positive $x$-direction in the region $x>q$, there is
no reflections at the interface at $x=q$.

In the domain before the barrier, $x < -q$, and from the continuity
conditions for $W,\ W'$ and $W''$ at $x=-q$ and from Eq.\
(\ref{eq:ansatz}), the Schr\"{o}dinger wave function is presented
as [\ref{bib:f95}]

\begin{equation}
\psi = \left( \frac{{\cal A}}{\hbar k}\right)^{1/2} \exp[i
\arctan({\cal B})],\  \   x<-q
\label{eq:ipsi}
\end{equation}

\noindent where

\begin{eqnarray}
{\cal A} & = &  \cosh ^2(-2\kappa q) + \frac{1}{2} \left(
\frac{\kappa }{k} + \frac{k}{\kappa } \right) \sinh (-4\kappa q)
\sin [2k(x+q)] \nonumber \\ [.08in] 
&  & \ \ \ \ + \sinh ^2(-2\kappa q) \left[ \left( \frac{\kappa }{k}
\sin [k(x+q)] \right) ^2 + \left( \frac{k}{\kappa } \cos[k(x+q)]
\right) ^2 \right]
\label{eq:calA}   
\end{eqnarray}

\noindent and

\begin{equation}
{\cal B} = \frac{\frac{k}{\kappa } \sinh (-2\kappa q) \cos [k(x+q)]
+ \cosh (-2\kappa q) \sin [k(x+q)]}{\cosh (-2\kappa q) \cos
[k(x+q)] + \frac{\kappa }{k} \sinh (-2\kappa q) \sin [k(x+q)]}. 
\label{eq:calB}
\end{equation}   

\noindent The Schr\"{o}dinger wave function, as represented by
Eqs.\ (\ref{eq:fpsi}) and (\ref{eq:ipsi}), has a continuous
logarithmic derivative across the barrier interface at $x=-q$. 
Similar to the situation at the barrier interface at $x=q$, Eqs.\
(\ref{eq:fpsi}) and (\ref{eq:ipsi}) manifest a wave with compound
spatial modulation of phase and amplitude for $x<q$ that progresses in the 
positive $x$-direction.  This wave with compound spatial modulation has a
continuous logarithmic derivative at $x=-q$, so there is no
refection of this wave at the barrier interface $x=q$.

The Schr\"{o}dinger wave function, as represented by Eqs.\ 
(\ref{eq:tpsi})--(\ref{eq:ipsi}), manifests a running wave
progressing in the positive $x$-direction everywhere.  Nowhere is
there any reflection of this running wave.  This Schr\"{o}dinger
wave function is an eigenfunction with eigenvalue energy $E$  for
the given rectangular barrier.  Hence, this eigenfunction
represents a particle with sub-barrier energy that tunnels through
the barrier with certainty.

Only recently did physicists recognize that eigenfunctions for a
constant potential could be wave functions with compound spatial
modulation in amplitude and wavenumber [\ref{bib:f94}].  However, mathematicians 
knew it all along, cf.\ Appendix A.  While one
could confirm that the wave function represented by Eqs.\
(\ref{eq:tpsi}) through (\ref{eq:ipsi}) is an eigenfunction by
brute force by substituting this wave function into the
Schr\"{o}dinger equation, we suggest referring to Ref.\ \ref{bib:f95} where the 
wave function representations, Eq.\ (\ref{eq:fpsi}) has been resolved into its 
customary hyperbolic components inside the barrier by

\begin{equation}
\psi = \frac{1}{(\hbar k)^{1/2}}\Bigl(\cosh [\kappa (x-q)] + i(k/\kappa )
\sinh [\kappa (x-q)]\Bigr), -q \le x \le q
\label{eq:fpsir}
\end{equation}

\noindent and where Eq.\ (\ref{eq:ipsi}) has been resolved into the customary 
incident and reflected unmodulated plane-wave components before the barrier by

\begin{eqnarray}
\psi \ & = & \ \overbrace{(\hbar k)^{-1/2} \left[ \cosh (-2\kappa q) + 
\frac{i}{2} \left( \frac{k}{\kappa } - \frac{k}{\kappa }\right) \sinh(-2\kappa 
q)\right] \exp[ik(x+q)]}^{\mbox{customary unmodulated incident plane wave}} 
\nonumber \\
& & \ \ \ \ \  + \ \underbrace{(\hbar k)^{-1/2}\frac{i}{2}\left( \frac{k}{\kappa 
}-\frac{\kappa }{k}\right) \sinh(-2\kappa q) \exp[-ik(x+q)]}_{\mbox{customary 
unmodulated reflected plane wave}}, \ \ \ \ x<q.  
\label{eq:ipsir}
\end{eqnarray}

\noindent Hence, Eqs. (\ref{eq:fpsi}) and (\ref{eq:ipsi}) manifest synthesized 
waves in and before the barrier respectively.

We note that the synthetic incident wave, Eq. (\ref{eq:ipsi}), has spectral 
components traveling in both the positive and negative $x$ directions.  Any 
concern that the synthetic wave, Eq. (\ref{eq:ipsi}), would spontaneously split 
apart is put to rest in Appendix A.

\subparagraph{2.3.  Bound States:}  The boundary value problem is not so
simple [\ref{bib:f82b},\ref{bib:f86c}].  The solutions for boundary value
problem, if they exist at all, need not be unique.  As is well
known for bound states, solutions for the Schr\"{o}dinger wave
function do exist for the energy eigenvalues.  Not as well known,
solutions for Hamilton's characteristic function for the trajectory
representation of quantum mechanics exist for bound states if the
action variable, $J$, is quantized [\ref{bib:mi30}], that is

\begin{equation}
J = \oint W'\, dx = nh, \ \ n=1,2,3,\cdots .
\label{eq:ave}
\end{equation}  

\noindent The action variable is independent of the set of
coefficients $(a,b,c)$ by the theory of complex variables [\ref{bib:f86c}].  The 
set of coefficients $(a,b,c)$ only posits the
singularities (poles) and terminal points of the Riemann sheets.
The set of coefficients $(a,b,c)$ does not effect the number of
poles or Riemann sheets.  

Specifically, we consider the bound state problem where  $\psi
\rightarrow 0$  as  $x \rightarrow \pm \infty$.  These are the
bound state eigenfunctions which are unique.  While the
Schr\"{o}dinger wave function is unique for bound states, the
conjugate momentum is not [\ref{bib:f86c},\ref{bib:f96b}].  In the generalized
Hamilton-Jacobi representation of quantum mechanics, the boundary
conditions for bound motion manifest a phase-space trajectory with
turning points at  $x = \pm \infty$.  This is accomplished by $W'
\rightarrow 0$ as $x \rightarrow \pm \infty$.  However, the
generalized Hamilton-Jacobi equation for the bound states is a
nonlinear differential equation that has critical (singular) points
at the very location where the boundary values are applied, i.e., 
$x = \pm \infty$.  By Eq.\ (\ref{eq:cme}),  $W' \rightarrow 0$  as
$x \rightarrow \pm \infty$  because at least one of the independent
solutions, $\phi $ or $\theta $, of the Schr\"{o}dinger equation must
be unbound as  $x \rightarrow \pm \infty$.  As the coefficients
satisfy  $a,b > 0$  and  $ab > c^2/4$,  the conjugate momentum
exhibits a node as $x \rightarrow \pm \infty$ for all permitted
values of $a$, $b$, and $c$ [\ref{bib:f86c}].  Hence, the boundary
values, $W'(x=\pm \infty ) = 0$, for Eq.\ (\ref{eq:hje}) permit 
non-unique phase-space trajectories for $W'$ for energy eigenvalues
or quantized action variables.  Likewise, the trajectories in
configuration space are not unique for the energy eigenvalue as the
equation of motion, $t-\tau = \partial W/\partial E$, specifies a
trajectory dependent upon the coefficients $a$, $b$ and $c$. 

\subparagraph{2.4.  Microstates:} The non-unique trajectories in phase
space and configuration space manifest microstates of the
Schr\"{o}dinger wave function [\ref{bib:f86c},\ref{bib:f96b}].  For bound
states in one dimension, the time-independent Schr\"{o}dinger wave
function may be real except for an inconsequential phase factor. 
Bound states have the boundary values that $\psi (x=\pm \infty ) =
0$.  Let us choose $\phi $ to be the bound solution. Then $\psi =
\alpha \phi + \beta \theta $ where $\alpha $ and $\beta $ are
coefficients. The Schr\"{o}dinger wave function for bound states
can be represented by [\ref{bib:f94}] 

\begin{eqnarray}
\psi & = & \frac{(2m)^{1/4} \cos(W/\hbar )}{(W')^{1/2}[a-
c^2/(4b)]^{1/2}} \nonumber \\
& = & \frac{(a\phi ^2+b\theta ^2+c\phi \theta )^{1/2}}{[a-
c^2/(4b)]^{1/2}} \cos \left[\arctan \left(\frac{b(\theta /\phi ) +
c/2}{(ab-c^2/4)^{1/2}}\right)\right] = \phi . 
\label{eq:cosansatz}
\end{eqnarray}

\noindent Thus,  $\alpha = 1$  and  $\beta = 0$  for all permitted
values of the set $(a,b,c)$.  Each of these non-unique trajectories
of energy $E$ manifests a microstate of the Schr\"{o}dinger wave
function for the bound state.  These microstates of energy $E$ are
specified by the set $(a,b,c)$.  See Ref.\ \ref{bib:f96b} for an example.

The existence of microstates is a counter-example refuting the
assertion of the Copenhagen interpretation that the Schr\"{o}dinger
wave function be the exhaustive description of nonrelativistic
quantum phenomena. 

Historically, others including Ballinger and March [\ref{bib:bm54}], Light and 
Yuan [\ref{bib:ly73}], and Korsch [\ref{bib:ko85}] had noted that the 
bound-state solution to Eq.\
(\ref{eq:hje}), or its equivalent by transformation, was arbitrary. 
(There may be others of whom I am unaware.)  These investigators enjoyed freedom 
in choosing the coefficients $(a,b,c)$ or their equivalents.  These 
investigators choose the particular solution that
rendered well behaved results for the density of states close to
WKB values [\ref{bib:ko85}] or gave good fits to extended Thomas-Fermi
approximations [\ref{bib:bm54},\ref{bib:ly73}].
Ballinger and March [\ref{bib:bm54}] and Korsch [\ref{bib:ko85}] acknowledge 
that their
choices of the particular solution, while fitting the work at hand,
could not be justified from quantum principles.   
 
\subparagraph{2.5.  Classical Limit, Loss of Information, Heisenberg Uncertainty 
and Residual Indeterminacy:}  For the classical limit $(\hbar \to 0)$, the 
QSHJE, a third-order non-linear differential equations, reduces to the classical 
stationary Hamilton-Jacobi equation (CSHJE), a first-order nonlinear 
differential equation.  Reducing the order in turn reduces the set of initial 
values necessary and sufficient to establish unique solution.  Hence, less 
information is necessary to solve the CSHJE than the QSHJE.  For the CSHJE, 
simultaneous knowledge of momentum and position specifies the energy and the 
trajectory.  While momentum and position form a sufficient set of initial 
conditions for classical mechanics, quantum mechanics also needs the higher 
order derivatives $W''$ and $W'''$ [\ref{bib:f84}].  The Heisenberg uncertainty 
principle alleges uncertainty in such simultaneous knowledge implying that 
trajectories do not exist at the quantum level.  This is premature as momentum 
and position form only a subset smaller than the set of initial conditions 
necessary and sufficient to solve the QSHJE [\ref{bib:f00a}].  

We note that this loss of information differs with the recent proposal of 
't~Hooft [\ref{bib:ho99}] that quantization results from the loss of information 
about ``primordial" trajectories of continuous energy.  No dissipation of 
information happens in the trajectory representation when going to the classical 
limit, but rather this loss of information induces an indeterminacy.  

As $\hbar \to 0$, we can test Planck's correspondence principle as to whether 
quantum mechanics goes to classical mechanics.  In the trajectory 
representation, the equation of motion for a free particle (i.e., $V=0$) can be 
expressed as [\ref{bib:f00a}]

\begin{equation}
t-t_o=\frac{(ab-c^2/4)^{1/2}(2m/E)^{1/2}x}{a+b+(a^2-2ab+b^2+c^2)^
{1/2}\cos \{2(2mE)^{1/2}x/\hbar +\cot ^{-1}[c/(a-b)]\}}.
\label{eq:qeom}
\end{equation}
     
\noindent In the limit $\hbar \to 0$, the cosine term in the denominator of Eq.\ 
(\ref{eq:qeom}) fluctuates with an infinitesimal short wavelength.  For the 
particular case, $a=b$ and $c=0$, Plank's correspondence principle holds for 
Eq.\ (\ref{eq:qeom}).  On the other hand for $a \ne b$ or $c \ne 0$, the cosine 
term becomes indefinite in the classical limit.  This leads to a residual 
indeterminacy in the classical limit.  Thus, Planck's correspondence principle 
does not hold in general.  This is consistent with the findings of Faraggi and 
Matone [\ref{bib:fm98a}--\ref{bib:fm00}] that the equivalence principle does not 
hold for classical mechanics [\ref{bib:f00a}]. It has also been shown elsewhere 
[\ref{bib:f00a}] that quantum mechanics does not reduce to statistical mechanics 
for $\hbar \to 0$.

Note that residual indeterminacy and the Heisenberg uncertainty principle 
differ:  the former exists for $\hbar \to 0$; the latter, for $\hbar $ finite 
[\ref{bib:f00a}].  Furthermore, Heisenberg uncertainty exists in the $[x,p]$ 
domain (where p is momentum) as the Hamiltonian operates in the $[x,p]$ domain.  
But the trajectory representation, through a canonical transformation to its 
Hamilton-Jacobi formulation, operates in the $[x,t]$ domain [\ref{bib:pa90}].  
Residual indeterminacy of the trajectory representation is in the $[x,t]$ 
domain, cf. Eq.\ (\ref{eq:qeom}).

In closing this subsection, we note that $\hbar $ remains finite and is very 
small.  Here, we treated $\hbar $ hypothetically as an independent variable to 
show even in the $\lim _{\hbar \to 0}$, quantum trajectories do not generally 
reduce to classical trajectories.       
    
\subparagraph{2.6.  Superluminality:}  The Aspect experiments deny local
reality [\ref{bib:agr82},\ref{bib:adr82}].  Yet the trajectories for bound 
states must penetrate
infinitely deep into the classically forbidden zone [\ref{bib:f82b}]. 
This infinitely long trip must be done in a finite period of time. 
Hence, superluminality follows.  This superluminality is a two-way 
superluminality.  An example that shows this is
given by Ref.\ \ref{bib:f00b}. 

Let us consider a particle traveling in a two-dimensional square-well duct.  The 
particle has a trajectory down the duct in the axial direction while vertexing 
at infinite turning points in the transverse direction.  The trajectory at these 
infinite turning points has been shown to be a cusp where velocity increases 
without bound and both legs of the cusp become tangent to the surface of 
Hamilton's characteristic function [\ref{bib:f00b}].  This manifests the extreme 
example that the trajectory is not generally orthogonal to the $W$-surface.  

Our trajectories incorporate reality by precept.  The underlying
generalized Hamilton-Jacobi equation is a phenomenological
equation.  Therefore, we find that since the trajectories have
reality inherently, they must describe a nonlocal reality where
phenomena violate Einstein separability.  Thus, the trajectory
representation renders a quantitative phenomenological description
that favors choosing quantum mechanics, albeit without the
Copenhagen interpretation thereof, in resolving the paradox between
quantum mechanics and Einstein separability that exists, for
example, in EPR experiments.

\paragraph{3. Different Predictions between Trajectories and Copenhagen:} 

\subparagraph{3.1.  Impulsive Perturbations:}  Floyd [\ref{bib:f99}] has shown 
that the trajectory and Copenhagen representations render different predictions 
for the first-order change in energy, $E_1$ due to a small, spatially symmetric 
perturbing impulse, $\lambda V(x)\delta(t)$, acting on the ground state of a 
infinitely deep, symmetric square well.  The different predictions are due to 
the different roles that causality plays in the trajectory and Copenhagen 
interpretations.  In the trajectory representation, $E_1$ is dependent upon the 
particular microstate, $(a,b,c)$.  This has been investigated under a Copenhagen 
epistemology even for the trajectory theory, where complete knowledge of the 
initial conditions for the trajectory as well as knowledge of the particular 
microstate are not necessary to show differences for an ensemble sufficiently 
large so that all microstates are individually well represented.  In the 
trajectory representation, the first-order change in energy, $E_1$, is due to 
the location of the particle in its trajectory when the impulse occurs.  The 
trajectory representation finds that the perturbing impulse, to first order, is 
as likely to do work on the particle as the particle is to do work perturbing 
system, cf. Eqs.\ (15) and (17)--(20) of Ref.\ \ref{bib:f99}.  Hence, the 
trajectory representation evaluates $\langle E_1 \rangle _{\mbox{\scriptsize 
average}} = 0$.  On the other hand, Copenhagen predicts $E_1$ to be finite as 
Copenhagen evaluates $E_1$ by the trace ground-state matrix element $\lambda 
V_{00} \delta(0)$ at the instant of impulse.  Due to spatial symmetry of the 
ground state and $V(x)$, $V_{00} \ne 0$.  

In an actual test, we do not need perturbing impulses, which were used for 
mathematical tractability.  A rapid perturbation whose duration is much shorter 
than the period of the unperturbed system would suffice [\ref{bib:f99}]. 

\subparagraph{3.2.  Overdetermination:}  For a square well duct, we have proposed a 
test where consistent overdetermination of the trajectory by a redundant set of
observed constants of the motion would be beyond the Copenhagen interpretation
[\ref{bib:f00b}].  The overdetermined set of constants of the motion should have 
a redundancy that is consistent with the particular trajectory.  On the other 
hand, Copenhagen would predict a complete lack of consistency among these 
observed constants of the motion as Copenhagen denies the existence of 
trajectories.  Such a test could be designed to be consistent with Copenhagen 
epistemology. 

\paragraph{4. Other Differences between Trajectories and Copenhagen:} As the 
trajectory exists by precept in the trajectory representation, there is no need 
for Copenhagen's collapse of the wave function.

The trajectory representation can describe an individual particle.  On the other 
hand, Copenhagen describes an ensemble of particles while only rendering 
probabilities for individual particles.

The trajectory representation renders microstates of the Schr\"{o}dinger wave
function for the bound state problem.  Each microstate by Eq.\
(\ref{eq:cosansatz}) is sufficient by itself to determine the
Schr\"{o}dinger wave function.  Thus, the existence of microstates
is a counter example refuting the Copenhagen assertion that the
Schr\"{o}dinger wave function be an exhaustive description of
nonrelativistic quantum phenomenon.

The trajectory representation is deterministic.  We can now
identify a trajectory and its corresponding Schr\"{o}dinger wave
function with sub-barrier energy that tunnels through the barrier
with certainty.  Hence, tunneling with certainty is a counter
example refuting Born's postulate of the Copenhagen interpretation
that attributes a probability amplitude to the Schr\"{o}dinger wave
function.

As the trajectory representation is deterministic and does not need $\psi $, 
much less  to assign a probability amplitude to it, the trajectory 
representation does not need a wave packet to describe or localize a
particle.  The equation of motion, Eq.\ (\ref{eq:eom}) for a
particle (monochromatic wave) has been shown to be consistent with
the group velocity of the wave packet [\ref{bib:f94}]. 

Normalization, as previously noted herein, is determined by the
nonlinearity of the generalized Hamilton-Jacobi equation for the
trajectory representation and for the Copenhagen interpretation by
the probability of finding the particle in space being unity. 

Though probability is not needed for tunneling through a barrier 
[\ref{bib:f95}], the trajectory interpretation for
tunneling is still consistent with the Schr\"{o}dinger
representation without the Copenhagen interpretation.  The incident
wave with compound spatial modulation of amplitude and phase for the trajectory 
representation, Eq.\ (\ref{eq:ipsir}), has only two spectral components which 
are the incident and reflected unmodulated waves of the Schr\"{o}dinger
representation [\ref{bib:f95}].

Trajectories differ with Feynman's path integrals in three ways.  First, 
trajectories employ a quantum Hamilton's characteristic function while a path 
integral is based upon a classical Hamilton's characteristic function.  Second, 
the quantum Hamilton's characteristic function is determined uniquely by the 
initial values of the QSHJE while path integrals are democratic summing over all 
possible classical paths to determine Feynman's amplitude.  While path integrals 
need an infinite number of constants of the motion even for a single particle in 
one dimension, motion in the trajectory representation for a finite number of 
particles in finite dimensions is always determined by only a finite number of 
constants of the motion.  Third, trajectories are well defined in classically 
forbidden regions where path integrals are not defined by precept. 

As previously noted in Section 2.5, the Heisenberg uncertainty principle shall 
remain premature as long as Copenhagen uses an insufficient subset of initial 
conditions $(x,p)$ to describe quantum phenomena.

Bohr's complementarity postulates that the wave-particle duality be resolved 
consistent with the measuring instrument's specific properties.  On the other 
hand, Faraggi and Matone [\ref{bib:fm98a}--\ref{bib:fm00}] have derived the 
QSHJE from an equivalence principle without evoking any axiomatic interpretation 
of the wave function.  Furthermore, Floyd [\ref{bib:f96b}] and Faraggi and 
Matone [\ref{bib:fm98a}--\ref{bib:fm00}] have shown that the QSHJE renders 
additional information beyond what can be gleaned from the Schr\"{o}dinger wave 
function alone. 

Anonymous referees of the Copenhagen school have had reservations
concerning the representation of the incident modulated wave as
represented by Eq.\ (\ref{eq:ipsi}) before the barrier.  They have
reported that compoundly modulated wave represented by Eq.\
(\ref{eq:ipsi}) is only a clever superposition of the incident and
reflected unmodulated plane waves.  They have concluded that
synthesizing a running wave with compound spatial modulation from
its spectral components is nonphysical because it would spontaneously split.  We 
have put these reservations to rest in Appendix A and Ref.\ \ref{bib:f95}.  By 
the superposition principle
of linear differential equations, the spectral components may be
used to synthesize a new pair of independent solutions with compound
modulations running in opposite directions. Likewise, an
unmodulated plane wave running in one direction can be synthesized
from two waves with compound modulation running in the opposite
directions for mappings under the superposition principle are
reversible.  

\paragraph{5. Trajectories vis-a-vis Bohmian mechanics:}  The
trajectory representation differs with Bohmian representation 
[\ref{bib:bo52},\ref{bib:bh87}] in many ways despite both
representations being based on equivalent generalized 
Hamilton-Jacobi equations.  We describe the various differences
between the two representations in this section.  These
differences may not necessarily be independent of each other.

First, the two representations have different equations of motion.
The Hamilton-Jacobi transformation equation, Eq.\ (\ref{eq:eom}),
are the equations of motion for the trajectory representation. 
Meanwhile, Bohmian mechanics eschews solving the Hamilton-Jacobi
equation for a generator of the motion, but instead assumes that
the conjugate momentum be the mechanical momentum, $m\dot{x}$,
which could be integrated to render the trajectory.  But the
conjugate momentum is not the mechanical momentum as already shown
by Floyd [\ref{bib:f82b},\ref{bib:f94}], Faraggi and Matone [\ref{bib:fm00}] and 
Carroll [\ref{bib:ca99}].  Recently, Brown and Hiley [\ref{bib:bh00}] had stated 
that prior associating momentum in Bohmian mechanics with $W'$ by appealing to 
classical canonical theory was a ``backward step'' and ``totally unnecessary''.  
Brown and Hiley still do not advocate solving the QSHJE for $W$.  Rather, they 
now advocate that $W'$ be a ``beable" momentum and $\dot{x}$ be given by the 
probability current divided by the square of the probability amplitude.   

Bohmian mechanics considers $\psi $ to form a field, a quantum
field that fundamentally effects the quantum particle.  The
trajectory representation considers the Schr\"{o}dinger equation to
be only a phenomenological equation where $\psi $ does not
represent a field.  To date, no one has ever measured such a $\psi
$-field.

Bohmian mechanics postulates a quantum potential, $Q$, in addition
to the standard potential, that renders a quantum force
proportional to $-\nabla Q$. \{Bohm's quantum potential in one dimension appears 
in the QSHJE as the negative of the term containing the Schwarzian derivative or 
the right side of Eq.\ (\ref{eq:hje}), i.e., $Q=[\hbar ^2/(4m)]\langle W;x 
\rangle \}$.  But this quantum potential is inherently dependent upon $E$.  By 
the QSHJE, $Q$ is also dependent upon the microstate $(a,b,c)$ of a given 
eigenvalue energy $E$ because

\[
Q = E - V  + (a \phi ^2 + b \theta ^2 + c \phi \theta )^{-2}.
\]

\noindent Therefore, $Q$ as a function is path dependent and cannot be a 
conservative potential.  Consequently, $-\nabla Q$ does not generally render a 
force. The average energy associated with $Q$ or the Schwarzian derivative term 
of the QSHJE in the classical limit $(\hbar \to 0)$ for the free particle 
$(V=0)$ is dependent upon the microstate as specified by $(a,b,c)$ and is given 
by [\ref{bib:f00a}] 

\begin{equation}
\Bigl \langle \lim_{\hbar \to 0} Q \Bigr \rangle _{\mbox{\scriptsize average}} =
E\left(1-\frac{(a+b)/2}{(ab-c^2/4)^{1/2}}\right) =
-\frac{\mbox{variance\ of\ }{\textstyle \lim_{\hbar \to 0}}W_x}{2m}
\le 0.
\label{eq:bqp}
\end{equation}

\noindent So the average energy, in the classical limit of Bohm's quantum 
potential, $Q$, is proportional to the negative of the variance of the classical 
limit of the conjugate momentum.  The quantum potential is a function of the 
particular microstate and may be finite even in the classical limit as shown by 
Eq.\ (\ref{eq:bqp}).  Nothing herein implies that Eq.\ (\ref{eq:bqp}) is 
general.  Others cases have not been examined.

While Bohmian mechanics postulates pilot waves to guide the particle, the 
trajectory representation does not need any such waves.

Bohmian mechanics uses an {\it ansatz} that contains an exponential
with imaginary  arguments.  The Bohmian {\it ansatz} in one
dimension is $\psi = (W')^{-1/2} \exp (iW/\hbar )$, the same as Eq. 
(\ref{eq:ansatz}). Anonymous referees of the Bohm school have expressed 
reservations regarding the validity of trigonometric {\it ans\"{a}tze}. Herein, 
we have presented, without using any particular {\it ansatz}, the reversible 
relationship between the generalized Hamilton-Jacobi
equation, Eq.\ (\ref{eq:hje}), to the Schr\"{o}dinger equation by
Eq.\ (\ref{eq:hjse}).    As Eq.\ (\ref{eq:hjse}) is valid for any set $(\phi 
,\theta )$, other {\it ans\"{a}tze}  of the form $\psi = (W')^{-1/2} [A \exp 
(iW/\hbar) + B \exp (-iW/\hbar)]$, where $A,B$ are arbitrary, are acceptable 
[\ref{bib:f82b},\ref{bib:fm00}].  When $|A|=|B|$, then the {\it ansatz} becomes 
trigonometric.  In the past, the trajectory representation had properly used 
other {\it ans\"{a}tze} that were trigonometric in nature such as Eq.\ 
(\ref{eq:cosansatz}). For completeness, Bohm's {\it ansatz} has significantly 
more versatility than first apparent if $|A| \ne |B|$.  Consider  

\begin{eqnarray*}
\psi & = & (W')^{-1/2}\frac{A+B}{A-B} [A \exp (iW/\hbar) + B \exp (-iW/\hbar)] 
\\
     & = & (W')^{-1/2}\frac{A+B}{A-B}[(A+B)^2\cos ^2(iW/\hbar) + (A-B)^2\sin 
^2(iW/\hbar)]^{1/2} \exp \left[ \frac{i}{\hbar}\arctan \left( \frac{A-
B}{A+B}\tan (W)\right) \right] \\
     & = & (\widetilde{W}')^{-1/2} \exp (i\widetilde{W}/\hbar ), \ \ \ |A| \ne 
|B|
\end{eqnarray*}

\noindent where 

\[
\widetilde{W} = \arctan \left(\frac{A-B}{A+B} \tan (W)\right).
\]

\noindent So, we have returned to Bohm's one-dimensional {\it ansatz} with a new 
Hamilton's characteristic function $\widetilde{W}$ for $|A| \ne |B|$.  This {\it 
ansatz} is reminiscent of the modulated wave that we presented in Eqs.\ 
(\ref{eq:ansatz}) and (\ref{eq:ipsi}).
   	
Bohmian mechanics asserts that particles could never reach a point
where the Schr\"{o}dinger wave function vanishes.  On the other
hand, trajectories have been shown to pass through nulls of $\psi $ 
[\ref{bib:f82b},\ref{bib:fm00}].  Furthermore, the conjugate momentum is finite 
at these nulls by Eq.\ (\ref{eq:cme}) as $\phi $ and $\theta $ cannot
be both zero at the same point for they are independent solutions
of a linear differential equation of second order.

Bohmian mechanics asserts that bound-state particles should have zero
velocity because the spatial part of the bound-state wave function can be
expressed by a real function.  On the other hand, the generalized
Hamilton-Jacobi equation, Eq.\ (\ref{eq:hje}) is still applicable
for bound states in the trajectory representation.  For bound
states, the trajectories form orbits whose action variables are
quantized according to Eq.\ (\ref{eq:ave}).
  
Bohmian mechanics asserts that a particle should follow a path normal
to the surfaces of constant $W$.  On the other hand, our
trajectories, when computed in higher dimensions, are not generally normal to 
the surfaces of constant $W$ [\ref{bib:f00b},\ref{bib:f94}]. 
In higher dimensions, the trajectories are determined by the 
Hamilton-Jacobi transformation equations for constant coordinates
(Jacobi's theorem) rather than by $\nabla W$.    

Bohmian mechanics asserts that the possible Bohmian trajectories
for a particular particle should not cross.  Rather, Bohmian
trajectories are channeled and follow hydrodynamic-like flow
lines.  On the other hand, the trajectory representation describes
trajectories that not only can cross but can also form caustics as
shown elsewhere in an analogous, but applicable acoustic two-dimensional duct  
[\ref{bib:f86a}].  We note that the Schr\"{o}dinger eaquation and the separated 
acoustic wave equations are both Helmholtz equations.
   
The two representations differ epistemologically whether
probability is needed.  The trajectory representation is
deterministic.  Bohmian mechanics purports to be stochastic and
consistent with Born's probability amplitude [\ref{bib:bh87}]. 
In one dimension, Bohmian mechanics introduces stochasticity, by
assigning  a position, $\chi $, of the particle as a separate
variable from the argument, $x$, of the Schr\"{o}dinger wave
function, $\psi $.  In other words, Bohmian mechanics introduces
stochasticity by assuming different initial positions of the
particle within the initial wave packet for the probability
amplitude of the particle.  The particle position, $\chi $, would
be a stochastic variable.  From Bell [\ref{bib:be87}],  the argument $x$ of
$\psi $ could be treated as the ``hidden'' variable instead of
$\chi $.  We note that this additional variable, $\chi $, is
extraneous for consistency with the Schr\"{o}dinger equation [\ref{bib:f94}].

Let us consider three dimensions in this paragraph to examine the familiar 
stationary auxiliary equation 

\begin{equation}
\nabla \cdot (R^2\nabla W) = 0
\label{eq:continuity}
\end{equation}

\noindent to the three-dimensional QSHJE. Bohm and Hiley [\ref{bib:bh87}] 
identify $R$ as  a probability amplitude and Eq. (\ref{eq:continuity}) as the 
continuity equation conserving probability.  Bertoldi, Faraggi and Matone 
[\ref{bib:bfm99}] only require that $R$ satisfy Eq.\ (\ref{eq:continuity}) 
nontrivially.  Hence, $R^2\nabla W$ must be divergenceless.  The trajectory 
representation can now show a non-probabilistic interpretation of $R^2\nabla W$.  
Let us consider a case for which the stationary Bohm's {\it ansatz}, $\psi = R 
\exp(iW/\hbar)$, is applicable.  Bohm used [\ref{bib:bo52}]  

\[
R^2={\cal U}^2 + {\cal V}^2 \ \ \ \mbox{and} \ \ \ W=\hbar \arctan({\cal 
V}/{\cal U})
\]

\noindent where ${\cal U} = \Re (\psi ) = R \cos (W/\hbar )$ and ${\cal W} = \Im 
(\psi ) = R \sin (W/\hbar )$.  Hence, by the superposition principle, ${\cal U}$ 
and ${\cal V}$ are a pair of solutions, not necessarily independent, to the 
stationary Schr\"{o}dinger equation.  (If ${\cal U}$ and ${\cal V}$ are not 
independent, then $W$ is a constant and $\psi $ is real except for a constant 
phase factor.) Upon substituting ${\cal U}$ and ${\cal V}$ into Eq.\ 
(\ref{eq:continuity}), we get as an intermediate step

\[
R^2\nabla W = {\cal U}\nabla {\cal V} - {\cal V}\nabla {\cal U},
\]

\noindent which is like a three-dimensional Wronskian.  Again, we do not need 
this Wronskian analogy to be a constant, just divergenceless.  The divergence of 
$R^2\nabla W$ is

\[
\nabla \cdot (R^2\nabla W) = \nabla {\cal U} \nabla {\cal V}(1-1) + 
\frac{2m}{\hbar ^2}(E-V){\cal U}{\cal V}(1-1) = 0.
\]

\noindent Indeed $R^2\nabla W$ is divergenceless.  Thus, the trajectory 
representation finds that the auxiliary equation contains a three-dimensional 
Wronskian analogy that satisfies Eq.\ (\ref{eq:continuity}) without any need for 
evoking a probability amplitude. 

Bohm had expressed concerns regarding the initial distributions of
particles.  Bohm [\ref{bib:bo52}] had alleged that in the duration that
nonequilibrium probability densities exist in his stochastic 
representation, the  usual formulation of quantum mechanics would
have insoluble difficulties.  The trajectory representation has
shown that the initial conditions of nonlocal hidden variable may
be arbitrary and still be consistent with the Schr\"{o}dinger
representation [\ref{bib:f84}].

Stochastic Bohmian mechanics, like the Copenhagen interpretation,
uses a wave packet to describe the motion of the of the associated
$\psi $-field.  As previously described herein, the deterministic
trajectory needs neither waves nor wave packets to describe or localize
particles.

Holland [\ref{bib:ho93}] reports that the Bohm's equation for particle
motion could be deduced from the Schr\"{o}dinger equation but the
process could not be reversed.  On the other hand, the development
of Eq.\ (\ref{eq:hjse}) is reversible.  The Schr\"{o}dinger
equation and the generalized Hamilton-Jacobi equation mutually
imply each other.

In application, the two representations differ regarding
tunneling.  Dewdney and Hiley [\ref{bib:dh82}] have used Bohmian mechanics to
investigate tunneling through a rectangular barrier by Gaussian
pulses.  While Dewdney and Hiley assert consistency with the
Schr\"{o}dinger representation, they do not present any results in
closed form.  Rather, they present graphically an ensemble of
numerically computed trajectories for eye-ball integration to show
consistency with the Schr\"{o}dinger representation.  On the other
hand, our trajectory representation exhibits in closed form
consistency with the Schr\"{o}dinger representation (the unbound
wave function does not have microstates [\ref{bib:f96b}]).  In
addition, we note that every Bohmian trajectory that successfully
tunnels slows down while tunneling.  Hence, a particle following
any one of these Bohmian trajectories would slow down while
tunneling even though Steinberg et al [\ref{bib:st94}] have shown that the
peak of the associated wave packet speeds up while tunneling.  On
the other hand Floyd [\ref{bib:f00b},\ref{bib:f95}] has shown that trajectories 
that successfully tunnel speed up consistent with the findings of
Olkhovsky and Racami [\ref{bib:or92}] and Barton [\ref{bib:ba86}] and the 
finding of Hartmann [\ref{bib:ha62}] and Fletcher [\ref{bib:fl85}] for thick 
barriers.   

\bigskip

\renewcommand{\theequation}{A\arabic{equation}}
\setcounter{equation}{0}

\paragraph{Appendix A --- Inverse Mapping:}  In this Appendix we show that no 
particular set of independent solutions is privileged [\ref{bib:f95}].  The 
incident wave with compound spatial modulation of amplitude and phase, Eq. 
(\ref{eq:ipsi}), can be synthesized under the superposition principle from two 
spectral components running in opposite directions as shown by Eq. 
(\ref{eq:ipsir}).  Likewise, an unmodulated plane wave running in one direction 
can be synthesized from two waves with compound modulation running in opposite 
directions for mappings under the superposition principle are reversible.

As a heuristic example consider analyzing the unmodulated plane wave 
(eigenfunction for the free particle with energy $E$) into the solution set 
$(\zeta _+,\zeta _-)$ where

\[
\zeta _{\pm} = \left( \frac{{\cal A}}{\hbar k}\right)^{1/2} \exp[\pm i
\arctan({\cal B})]
\]

\noindent and where ${\cal A}$ and ${\cal B}$ have already been specified by 
Eqs.\ (\ref{eq:calA}) and (\ref{eq:calB}) respectively.

Hence, $\zeta _+$ and $\zeta _-$ are two modulated waves that run in opposite 
directions as there phases monotonically increase or decrease respectively with 
increasing $x$.  The customary incident and reflected unmodulated plane waves 
before the barrier are given respectively by [\ref{bib:f95}]

\begin{eqnarray}
& (\hbar k)^{-1/2} & \left[ \cosh (-2\kappa q) + \frac{i}{2} \left( 
\frac{k}{\kappa } - \frac{k}{\kappa }\right) \sinh(-2\kappa q)\right] 
\exp[ik(x+q)] \nonumber \\
&    & = \ \left[ \cosh ^2(-\kappa q) + \frac{1}{4} \left( \frac{k}{\kappa } - 
\frac{\kappa }{k} \right) ^2 \sinh ^2(-2\kappa q)\right] \zeta _+ \nonumber \\
&     & \ \ \ \ - \ \left[ \cosh (-\kappa q) + \frac{i}{2} \left( 
\frac{k}{\kappa } - \frac{\kappa }{k} \right) \sinh (-2\kappa q)\right] \left[ 
\frac{i}{2} \left( \frac{k}{\kappa } - \frac{\kappa }{k} \right) \sinh (-2\kappa 
q)\right] \zeta _-   
\label{eq:aincident}
\end{eqnarray}

\noindent and

\begin{eqnarray}
& \frac{{\displaystyle i/2}}{{\displaystyle (\hbar k)^{1/2}}} & \left( 
\frac{k}{\kappa } + \frac{\kappa }{k}\right) \sinh(-2\kappa q) \nonumber \\
&     & = \  -\frac{1}{4} \left( \frac{k}{\kappa } + \frac{\kappa }{k} \right) 
^2 \sinh ^2(-2\kappa q) \zeta _+ \nonumber \\
&     & \ \ \ \ + \left[ \cosh(-\kappa q) + \frac{i}{2} \left( \frac{k}{\kappa } 
- \frac{\kappa }{k} \right) \sinh (-2\kappa q)\right] \left[ \frac{i}{2} \left( 
\frac{k}{\kappa } - \frac{\kappa }{k} \right) \sinh (-2\kappa q)\right] \zeta _-
.   
\label{eq:areflected}
\end{eqnarray}

Equations (\ref{eq:aincident}) and (\ref{eq:areflected}) respectively map the 
customary incident unmodulated plane wave and the customary reflected 
unmodulated plane wave into the set $(\zeta _+,\zeta _-)$.  We have synthesized 
the customary incident and reflected unmodulated plane waves from two modulated 
waves, $(\zeta _+,\zeta _-)$, travelling in the opposite directions.  Hence, the 
superposition principle and its mappings are reversible.  If the customary 
unmodulated incident and reflected waves do not spontaneously split apart, than 
neither does the modulated incident wave.  If a pulse can be formed with 
unmodulated plane waves, so can the corresponding pulse be formed with modulated 
waves.  The set of unmodulated plane waves solutions to the time-independent 
Schr\"{o}dinger equation for a free particle is not privileged.
 
We note that Eq.\ (\ref{eq:aincident}), the customary unmodulated incident plane 
wave, and Eq.\ (\ref{eq:areflected}), the customary unmodulated reflected plane 
wave, sum to $\zeta _+$, which manifests the incident wave with compound spatial 
modulation, Eq.\ (\ref{eq:ipsi}), as expected.

\smallskip

\begin{center}
Acknowledgement
\end{center}

It is my pleasure to thank M. Matone for many discussions.  I also thank D. M. 
Appleby, G. Bertoldi, R.~Carroll, and A. E. Faraggi.

\smallskip

\subparagraph{References}

\small

\begin{enumerate}\itemsep -.02in
\item \label{bib:f} Floyd, E.\ R.: ``The Philosophy of the Trajectory 
Representation of Quantum Mechanics" to be published in R. Amoroso (ed.), {\it 
Proceedings of Vigier 2000 Symposium}, Kluwer Academic Publishers, Dordrecht. 
\item \label{bib:f82b} Floyd, E.\ R.: {\it Phys. Rev.} {\bf D 26} (1982), 
1339--1347.
\item \label{bib:f76a} Floyd, E.\ R.: {\it J.\ Math.\ Phys.}\ {\bf 17} (1976), 
880--884.
\item \label{bib:f79} Floyd, E.\ R.: {\it
J.\ Math.\ Phys.}\ {\bf 20} (1979), 83--85.
\item \label{bib:f82a} Floyd, E.\ R.: {\it Phys.\
Rev.}\ {\bf D 25} (1982), 1547--1551.
\item \label{bib:f76b} Floyd, E.\ R.:  {\it J.\ Acous.\ Soc.\ Am.}\ {\bf 60} 
(1976), 801--809.
\item \label{bib:f86b} Floyd, E.\ R.: {\it J.\ Acous.\ Soc.\ Am.}\ {\bf 80} 
(1986), 877--887.
\item \label{bib:f86c} Floyd, E.\ R.:  {\it Phys.\ Rev.}\ {\bf D 34} (1986), 
3246--3249.
\item \label{bib:fm98a} Faraggi, A.\ E.\ and Matone, M.:  {\it Phys.\ Lett.}\ 
{\bf A 249} (1998), 180--190, 
hep-th/9801033.
\item \label{bib:fm98b} Faraggi, A.\ E.\ and Matone, M.:  {\it Phys.\ Lett.}\ 
{\bf B 437} (1998), 369--380, hep-th/9711028.
\item \label{bib:fm98c} Faraggi, A.\ E.\ and Matone, M.: {\it Phys.\ Lett.}\ 
{\bf B 445} (1998), 77--81, hep-th/9809125.
\item \label{bib:fm99a} Faraggi, A.\ E.\ and Matone, M.:  {\it Phys.\ Lett.}\ 
{\bf B 445} (1999), 357--365, hep-th/9809126.
\item \label{bib:fm99b} Faraggi, A.\ E.\ and Matone, M.:  {\it Phys.\ Lett.}\ 
{\bf B 450} (1999), 34--40, 
hep-th/9705108.
\item \label{bib:fm00} Faraggi, A.\ E.\ and Matone, M.:  {\it Int.\ J.\ Mod.\ 
Phys.}\ {\bf A 15} (2000), 1869--2017, hep-th/9809127.
\item \label{bib:fa00a} Faraggi, A.\ E.\ (2000): ``Superstring phenomenology 
--- a personal perspective'', to appear in {\it Proceedings of Beyond the Desert 
99 --- Accelerator and non Accelerator Approaches}, hep-th/9910042
\item \label{bib:fa00b} Faraggi, A.\ E.\ (2000): ``Duality, equivalence, mass 
and the quest for the vacuum '', invited talk at PASCOS 99, hep-th/0003156.
\item \label{bib:ma00} Matone, M.\ (2000): ``Equivalence postulate and quantum 
origin of gravitation'', hep-th/0005274. 
\item \label{bib:bfm99} Bertoldi, G., A.\ E.\ Faraggi, A.\ E., and Matone, M. 
(1999): ``Equivalence principle, higher dimensional M\"{o}bius group and the 
hidden antisymmetric tensor of quantum mechanics'', in press {\it Clas. Quantum 
Grav.}, hep-th/99090201.
\item \label{bib:f99} Floyd, E.\ R.:   {\it Int.\ J.\ Mod.\ Phys.}\ {\bf A 14} 
(1999), 1111--1124, quant-ph/9708026.
\item \label{bib:f00b} Floyd, E.\ R.:  {\it Found.\ Phys.\ Lett.}\ {\bf 13} 
(2000), 235--251, quant-ph/9708007.
\item \label{bib:f88} Floyd, E.\ R.: {\it Int.\ J.\ Theo.\ Phys.}\ {\bf 27} 
(1988), 273--281.
\item \label{bib:f86a} Floyd, E.\ R.: {\it
J.\ Acous.\ Soc.\ Am.}\ {\bf 80} (1986), 1741--1747.
\item \label{bib:f94} Floyd, E.\ R.: {\it Phys.\ Essays} {\bf 7}, (1994) 
135--145.
\item \label{bib:f95} Floyd, E.\ R.: {\it An.\ Fond.\ Louis de Broglie} {\bf 20} 
(1995), 263--279.
\item \label{bib:f96a} Floyd, E.\ R.: {\it Phys.\ Lett.}\ {\bf A 214} (1996),  
259--265.
\item \label{bib:f84} Floyd, E.\ R.: {\it Phys.\ Rev.}\ {\bf D 29} (1984), 
1842--1844.
\item \label{bib:f96b} Floyd, E.\ R.: {\it Found.\ Phys.\ Lett.}\ {\bf 9} 
(1996), 489--497, 
quant-ph/9707051.
\item \label{bib:ca99} Carroll, R.: {\it J.\ Can.\ Phys.}\ {\bf 77} (1999), 
319--325, quant-ph/9903081.
\item \label{bib:mi30} Milne, W.\ E.:  {\it Phys.\ Rev.}\ {\bf 35} (1930), 
863--867.
\item \label{bib:bm54} Ballinger, R.\ A.\ and March, N.\ M.: {\it Proc. Phys.\ 
Soc.\ (London)} {\bf A 67} (1954), 
378--381.
\item \label{bib:ly73} Light, J.\ C.\ and Yuan, J.\ M.: {\it J.\ Chem.\ Phys.}\ 
{\bf 58} (1973), 660--671.
\item \label{bib:ko85} Korsch, H.\ J.: {\it Phys.\ Lett.}\ {\bf 109A} (1985), 
313--316.
\item \label{bib:f00a} Floyd, E.\ R.: {\it Int.\ J.\ Mod.\ Phys.}\ {\bf A 15} 
(2000), 1363--1378, quant-ph/9907092.
\item \label{bib:ho99} 't~Hooft, G.: {\it Class.\ Quantum Grav.}\ {\bf 16} 
(1999), 
3263--3279, gr-qc/9903084.
\item \label{bib:pa90} Park, D.: {\it Classical Dynamics and Its quantum 
Analogues}, 2nd ed., Springer-Verlag, New York 1990, p.\ 142.
\item \label{bib:agr82} Aspect, A., Grangier, P., and Roger G.: {\it
Phys.\ Rev.\ Lett.}\ {\bf 49} (1982), 91--94.
\item \label{bib:adr82} Aspect, A., Dalibard, J., and Roger, G.: {\it Phys.\ 
Rev.\ Lett.}\ {\bf 49} (1982), 1804--1807.
\item \label{bib:bo52} Bohm, D.:  {\it Phys.\
Rev.}\ {\bf 85} (1952), 166--179.
\item \label{bib:bh87} Bohm D.\ and Hiley, B.\ J.: {\it Phys.\ Rep.}\ {\bf 144} 
(1987), 323--348.
\item \label{bib:bh00} Brown, M. R. and Hiley, B. J.: ``Schr\"{o}dinger 
revisited: the role of Dirac's `standard' ket in the algebraic approach'', 
quant-ph/0005026.
\item \label{bib:be87} Bell, J.\ S.: {\it
Found. of Phys.} {\bf 12} (1982), 989--999; reprinted {\it
Speakable and Unspeakable in Quantum Mechanics} Cambridge, New
York, 1987, pp.\ 159--168.
\item \label{bib:ho93} Holland, P.\ R.: {\it The Quantum Theory of Motion},  
Cambridge U.\ Press, Cambridge, UK, 1993, p. 79.
\item \label{bib:dh82} Dewdney, C.\ and Hiley, B.\ J.: {\it Found.\ Phys.} 
(1982), {\bf 12}, 27--48.
\item \label{bib:st94} Steinberg, A.\ M., Kwiat, P.\ G.\ and Chiao, R.\ Y.:  
{\it Found. Phys.\ Lett.}\ {\bf 7} (1994), 223--237.
\item \label{bib:or92} Olkhovsky, V.\ S.\ and Racami, E.: {\it
Phys.\ Rep.}\ {\bf 214} (1992), 339--356.
\item \label{bib:ba86} Barton, G.: {\it An.\ Phys.\ (New York)} {\bf
166}, (1986), 322--363.
\item \label{bib:ha62} Hartmann, T.\ E.: {\it J.\ Appl.\ Phys.}\ {\bf 33} 
(1962), 3427--3433.
\item \label{bib:fl85} Fletcher, J.\ R.: {\it J.\ Phys.}\ {\bf C 18} (1985), 
L55--L59.

\end{enumerate}

\end{document}